\documentclass[prl,aps,twocolumn,showpacs,superscriptaddress]{revtex4}
\usepackage{graphicx,amssymb,amsmath,color,psfrag}
\usepackage{amsthm}
\usepackage{amsfonts}
\usepackage{algorithmic}
\usepackage{enumerate}
\usepackage{latexsym}

\input{tcilatex}
\begin{document}

\title{On vortex strength and beam propagation factor of fractional vortex beams}
\author{Jisen Wen}
\affiliation{Department of Physics, Zhejiang University, Hangzhou 310027, China}
\author{Li-Gang Wang}
\email{sxwlg@yahoo.com}
\affiliation{Department of Physics, Zhejiang University, Hangzhou 310027, China}
\author{Xihua Yang}
\affiliation{Department of Physics, Shanghai University, Shanghai 200444, China}
\author{Junxiang Zhang}
\affiliation{Department of Physics, Zhejiang University, Hangzhou 310027, China}
\author{Shi-Yao Zhu}
\affiliation{Department of Physics, Zhejiang University, Hangzhou 310027, China}

\begin{abstract}
Fractional vortex beams (FVBs) with non-integer topological charges attract
much attention due to unique features of propagations, but there still exist
different viewpoints on the change of their total vortex strength. Here we
have experimentally demonstrated the distribution and number of vortices
contained in FVBs at Fraunhofer diffraction region. We have verified that the
jumps of total vortex strength for FVBs happens only when non-integer
topological charge is before and after (but very close to) any even integer
number, which originates from two different mechanisms for generation and
movement of vortices on focal plane. Meanwhile, we have also measured the beam
propagation factor (BPF) of such FVBs, and have found that their BPF values
almost increase linearly in one component and oscillate increasingly in
another component. Our experimental results are in good agreement with
numerical results.

\end{abstract}
\date{\today }
\maketitle

\section{Introduction}

Vortex beams carrying phase singularities have attracted great research
interests. Usually, a phase singularity defines orbital angular momentum (OAM)
when its topological charge is integer \cite{Allen1992}. Vortex beams with
non-integer topological charges, called as fractional vortex beams (FVBs), and
sometimes also called as non-integer vortex beams, have interesting and unique
properties, such as alternating charge vortices \cite{Berry2004}, birth and
annihilation of vortices \cite{Leach2004}, topologically structured darkness
\cite{Alperin2017}, or "perfect" FVBs \cite{Georgiy2017}. FVBs have been used
to trap particles \cite{Tao2005} and their behaviors of phase singularities
are also adopted to demonstrate the Hilbert hotel paradox \cite{Gbur2016}. The
research of the FVBs has extended to acoustic wave \cite{Hong2015,Jia2018} and
the electron beams with fractional OAM states \cite{Bandyopadhyay2017}. The
fractional spiral phase also plays a crucial role in quantum information, for
example, fractional OAM momentum entanglement of two photons
\cite{Oemrawsingh2004,Oemrawsingh2005} and quantum digital spiral imaging
\cite{Chen2014}.

In fact, non-integer vortex beams are initially known as beams with mixed
screw-edge dislocations, which were discussed and generated experimentally by
computer-synthesized binary gratings in 1995 \cite{Basistiy1995}. Then in
2004, a new generation of FVBs was developed based on the production of
sub-harmonic diffraction by Basistiy \textit{et al } \cite{Basistiy2004}.
Moveover, ones found that the charge-1/2 vortex beam can be decomposed into a
series symmetric beams with different topological charges \cite{Basistiy2004}.
Later, Alperin \textit{et al.} used a stationary cylindrical lens to
quantitatively measure the OAM of FVBs, which is decomposed into intrinsic and
extrinsic \textquotedblleft modes\textquotedblright%
\ \cite{Alperin2016,Alperin2017}. However, due to complex propagation
properties of FVBs, the issue about the birth of vortices and the variation of
total vortex strength for FVBs has been extensively investigated but not been
solved yet.

Theoretical work by Berry predicts the birth of new vortex for FVBs
propagating in free space when topological charge is slightly greater than
half-integer \cite{Berry2004}, at which the vortex strength increases by unit.
This result was claimed to be confirmed by some experiments
\cite{Leach2004,Lee2004}. Recently, using large-angle holographic lithography
in photoresist with high resolution, Fang \textit{et al}. have visualized the
vortex birth and splitting of light fields induced by various non-integer
topological charges, and their result is consistent with Berry's prediction
\cite{Fang2017}. However, Jesus-Silva \textit{et al}. found that the birth of
vortices for FVBs at Fraunhofer zone happens at a small fraction which is much
smaller than half-integer topological charge. The vortex strength increases by
unit only at a number slightly larger than an integer. The result was also
confirmed in their experiment by using a triangular aperture to exam each
vortex at Fraunhfer zone \cite{Jesus2012}. However, the small fraction in
their results was hard to be determined, depending on beam parameters
\cite{Jesus2012}. Thus it is very necessary to further understand the dynamics
of vortices and the vortex strength of FVBs at far-field zone.

On the other hand, the beam width of FVBs increases as topological charge
increases at far-field zone, and their complex evolutions need a quantitative
description. As we know, the beam propagation factor (BPF) is a parameter of
great importance to characterize the global propagation feature of a light
beam. The BPF value proposed by Siegman is well defined by the second order
moment of intensity, which has intimate connection with far-field beam
divergent angle \cite{Siegman1990}. The BPF has been studied for many beams,
for example, Bessel-Gauss beams \cite{Borghi1997}, and elegant
Laguerre-Gaussian beams \cite{Porras2001}. Interestingly, theoretical works
have shown that the BPF value for integer Laguerre-Gaussian beams increases
linearly as topological charge increases \cite{Saghafi1998,Vega2007}. The
vortices can degrade beam quality, which has also been found by Ramee
\cite{Ramee2000}. Little attention has been paid to the BPF value of FVBs.
Naturally, it raises a question whether it still linearly increases as
topological charge increases for FVBs.

In this work, we have studied the vortex strength and BPF for FVBs. We have
reviewed and derived general expression for this kind of beams passing through
linear ABCD optical systems under paraxial approximation. More importantly, we
theoretically analyze and experimentally verify the vortex strength of FVBs at
Fraunhofer diffraction region. It shows the vortex strength increases only
when topological charge approaches an even number. This result is confirmed by
identifying all vortices in FVBs, using two methods: triangle aperture and
straight blade, and it is very different from previous understanding. To the
best of our knowledge, we have also demonstrated the first direct measurement
of the BPF value of FVBs with both x- and y- directions, and have explained
why there are different behaviors of the BPF values on these components.

\section{Theory}

First let us briefly review some main results about propagations of FVBs,
which are generated from light passing through non-integer spiral phase plates
\cite{Berry2004,Jesus2012,Gbur2016}. When a monochromatic plane-wave light
field initially propagates through an integer or non-integer phase plate at
$z=0$, then the initial electric field after this phase plate is simply given
by%
\begin{equation}
E_{\text{i}}(u,v,0)=\exp(i\alpha\phi),\label{Ein-p}%
\end{equation}
where $\alpha$ can be a positive, negative integer, or an arbitrary
non-integer number, and it is called as a topological charge number,
$\phi=\arctan(v/u)$ is the azimuthal angle, and $u,v$ are the transverse
rectangular coordinates at $z=0$. By neglecting the contribution of evanescent
waves and limiting to paraxial approximation, for the simplicity of the later
discussion, the propagations of the output light fields in linear ABCD optical
systems are described in terms of rectangular coordinates by the Collins
formula \cite{Collins1970,Zhao2000}
\begin{align}
E_{\text{o}}(x,y,z) &  \!=\!\frac{\exp(ikL)}{i\lambda B}\!\int\!\int
\!E_{\text{i}}(u,v,0)\exp\{\frac{ik}{2B}[A(u^{2}+v^{2})\nonumber\\
&  -2(xu+yv)+D(x^{2}+y^{2})]\}dudv,\label{Eout}%
\end{align}
where $A$, $B$, and $D$ are the elements of a $2\times2$ ray transfer matrix
$\left(
\begin{smallmatrix}
A & B\\
C & D
\end{smallmatrix}
\right)  $ describing a linear optical system, $L$ is the eikonal along the
propagation axis, and $k$ is the wave number of light. The above equation can
also be readily transferred into the cylindrical coordinate system, and the
result of the output field for $\alpha=n$ being an integer is expressed by%
\begin{align}
E_{\text{o,}n}(\rho,\varphi,z) &  =(-i)^{\frac{|n|}{2}}A^{-3/2}(\frac{\pi
k\rho^{2}}{8B})^{1/2}\exp\left[  i(kL+n\varphi)\right]  \nonumber\\
&  \times\exp\left[  \frac{ik\rho^{2}}{2B}\left(  D-\frac{1}{2A}\right)
\right]  \nonumber\\
&  \times\left[  J_{\frac{|n|-1}{2}}\left(  \frac{k\rho^{2}}{4AB}\right)
-iJ_{\frac{|n|+1}{2}}\left(  \frac{k\rho^{2}}{4AB}\right)  \right]
,\label{Eout2plane}%
\end{align}
where $\rho=(x^{2}+y^{2})^{1/2}$ is the radial coordinate, $\varphi
=\arctan(y/x)$ is the azimuthal angle, and $J_{n}(\cdot)$ is Bessel function
of first kind. This equation is the same with Eq. (6) in Ref. \cite{Berry2004}
when $A=1$, $B=z$, and $D=1$ for light propagation in free space.

In many practical situations including our experiment, the incident light
source is a fundamental Gaussian beam and then the initial field in Eq.
(\ref{Ein-p}) reads%
\begin{equation}
E_{\text{i}}(u,v,0)=\exp(-\frac{u^{2}+v^{2}}{w_{0}^{2}})\exp(i\alpha
\phi),\label{Ein-new}%
\end{equation}
where $w_{0}$ is the initial beam width. Thus Eq. (\ref{Eout2plane}) becomes
\begin{align}
E_{\text{o,}n}(\rho,\varphi,z)  & \!=\!\frac{(-i)^{|n|\!+\!1}z_{R}^{2}%
}{(B\!-\!iAz_{R})^{3/2}}(\frac{\pi\rho^{2}}{4Bw_{0}^{2}})^{1/2}\exp\left[
i(kL\!+\!n\varphi)\right]  \nonumber\\
& \!\times\!\exp\left[  \frac{ik\rho^{2}}{2B}\left(  D\!+\!\frac{iz_{R}%
}{2(B\!-\!iAz_{R})}\right)  \right]  \nonumber\\
& \times\left[  I_{\frac{|n|\!-\!1}{2}}\left(  \frac{z_{R}^{2}\rho^{2}%
/w_{0}^{2}}{2B(B\!-\!iAz_{R})}\right)  -I_{\frac{|n|\!+\!1}{2}}\left(
\frac{z_{R}^{2}\rho^{2}/w_{0}^{2}}{2B(B-iAz_{R})}\right)  \right]
,\label{Eout3Gaussian}%
\end{align}
where $z_{R}=kw_{0}^{2}/2$ is the Rayleigh distance of a Gaussian beam and
$I_{n}(\cdot)$ is modified Bessel function of first kind. When $w_{0}%
\rightarrow\infty$ (i.e., $z_{R}\rightarrow\infty$), it is clear that Eq.
(\ref{Eout3Gaussian}) can reduce into Eq. (\ref{Eout2plane}).

For $\alpha$ being a non-integer, following the method presented by Berry
\cite{Berry2004}, the non-integer phase distribution at the initial place can
be expanded into the Fourier series\setcounter{equation}{5}
\begin{equation}
\exp(i\alpha\phi)=\frac{\exp(i\pi\alpha)\sin(\pi\alpha)}{\pi}\sum
\limits_{n=-\infty}^{\infty}\frac{\exp(in\phi)}{\alpha-n}. \label{Expansion1}%
\end{equation}
Therefore the output fields for non-integer phase structures are given by
\begin{equation}
E_{\text{o,}\alpha}(\rho,\varphi,z)=\frac{\exp(i\pi\alpha)\sin(\pi\alpha)}%
{\pi}\sum\limits_{n=-\infty}^{\infty}\frac{E_{\text{o,}n}(\rho,\varphi
,z)}{\alpha-n}. \label{Expansion22}%
\end{equation}
Clearly, the results for the non-integer cases are involved with the
interference of all integer-vortex light fields. In Eq. (\ref{Expansion22}),
the function $E_{\text{o,}n}(\rho,\varphi,z)$ is to use Eq. (\ref{Eout2plane})
for a plane-wave light source while it is to use Eq. (\ref{Eout3Gaussian}) for
a finite Gaussian beam. Once the parameters of optical systems are known from
the ray transfer matrix $\left(
\begin{smallmatrix}
A & B\\
C & D
\end{smallmatrix}
\right)  $, the evolution of such integer and non-integer vortex light fields
can be analytically obtained. Meanwhile, one can also directly obtain the
propagation properties of FVBs from the numerical integration by substituting
Eq. (\ref{Ein-new}) into Eq. (\ref{Eout}). However, we emphasize that the
calculation accuracy by numerical integration or using Eq. (\ref{Expansion22})
are very important for obtaining the correct results, since it needs the
sufficient summations in Eq. (\ref{Expansion22}).

For light propagation in free space, the ray transfer matrix is simply given
by $\left(
\begin{smallmatrix}
A & B\\
C & D
\end{smallmatrix}
\right)  =\left(
\begin{smallmatrix}
1 & z\\
0 & 1
\end{smallmatrix}
\right)  $, then Eq.(\ref{Eout}) becomes $E_{\text{o}}(x,y,z)=\frac{\exp
(ikz)}{i\lambda z}\iint E_{\text{i}}(u,v,0)\exp\{\frac{ik}{2z}[(u^{2}%
+v^{2})-2(xu+yv)+(x^{2}+y^{2})]\}dudv$. In Refs.
\cite{Berry2004,Leach2004,Lee2004,Gbur2016}, it seems that the propagation
phenomena of FVBs are very complex in free space. In order to evaluate the
propagation behavior of such beams, one method is to measure their BPF (also
called as beam quality), which establishes the relation between the near-field
beam size and the far-field beam spread. Although Eqs. (\ref{Eout2plane}) and
(\ref{Eout3Gaussian}) are expressed in cylindrical coordinate systems, the
field distributions have no cylindrical symmetry in both the near-field and
far-field regions. Thus it is more convenient to discuss the BPF within
rectangular coordinates. The BPF values in both the $x$ and $y$ directions are
defined as \cite{Siegman1990},
\begin{subequations}
\begin{align}
M_{x}^{2}=4\pi w_{x}\Delta\theta_{x},\label{M2x}\\
M_{y}^{2}=4\pi w_{y}\Delta\theta_{y}, \label{M2y}%
\end{align}
where $w_{x}$ and $w_{y}$\ ($\Delta\theta_{x}$ and $\Delta\theta_{y}$) are the
square roots of the minimum spatial variances (the spatial-frequency
variances) in $x$ and $y$ directions, respectively, in terms of the
second-order moments associated with the intensities at initial plane
(far-field zone). Here $w_{q}$ ($q=x,y$) at the initial plane is calculated
by
\end{subequations}
\begin{equation}
w_{q}^{2}=\frac{\iint(q-\bar{q})^{2}|E_{\text{i}}(u,v,0)|^{2}dudv}%
{\iint|E_{\text{i}}(u,v,0)|^{2}dudv}. \label{Wq}%
\end{equation}
where $\bar{q}$ is the center of beam profile along this $q$ direction. In our
experiment, we should record the intensity distributions at $z=0$ (i.e., at
the plane of spatial light modulator, SLM), then $w_{q}$ can be obtained from
its intensity distribution. Theoretically, we have $w_{x}=w_{y}=w_{0}/2$ by
substituting Eq. (\ref{Ein-new}) into Eq. (\ref{Wq}).

\begin{figure}[htbp]
\centering
\includegraphics[width=8.4cm]{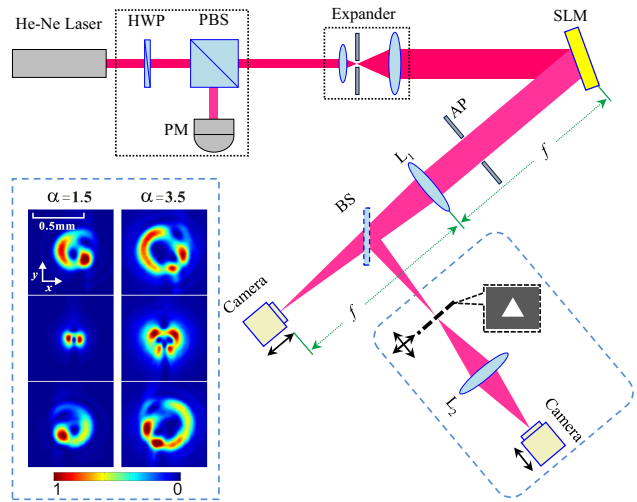}
\caption{(Color) Experimental setup for
measuring intensity distributions of FVBs near the focal plane of a 2-$f$ lens
system. When a beam splitter (BS) is inserted after the lens L$_{1}$, the
reflected light is used for verifying their topological charges by using a
triangle aperture placed on a two-axis translation stage. Left insert figure
is experimental intensity measurement of FVBs with $\alpha=1.5$ and 3.5,
respectively, locating at 40 mm before (top), at (middle), and at 40 mm after
(bottom) the focal plane. The lens L$_{1}$ has a focal length $f=300$ mm.
Other notations are: HWP, half-wave plate; PBS, polarized beam splitter; PM,
power meter; SLM, spatial light modulator; AP, Aperture. }%
\label{Fig1}%
\end{figure}

Meanwhile, in order to obtain the spatial-frequency variances $(\Delta
\theta_{q})^{2}$ at far-field zone, here a 2-$f$ focusing system with a focal
length $f$ (as shown in Fig. 1) is employed and enables us to measure the
far-field intensity distribution at its focal plane. The ray transfer matrix
of such a 2-$f$ focusing system is given by $\left(
\begin{smallmatrix}
A & B\\
C & D
\end{smallmatrix}
\right)  =\left(
\begin{smallmatrix}
0 & f\\
-1/f & 0
\end{smallmatrix}
\right)  $ at the focal plane, therefore the output field at the focal plane
is expressed by $E_{\text{o}}(x,y,f)=\frac{\exp(2ikf)}{i\lambda f}\iint
E_{\text{i}}(u,v,0)\exp\left[  -\frac{ik}{f}(xu+yv)\right]  dudv$. This
expression is very similar to the Fraunhofer diffraction equation of the far
field. According to the definition of spatial-frequency variables, $\theta
_{q}=q/(\lambda f)$, the field profile at focal plane changes into:
$E_{\text{o}}(\theta_{x},\theta_{y},f)=\frac{\exp(2ikf)}{i\lambda f}\iint
E_{\text{i}}(u,v,0)\exp\left[  -i2\pi(\theta_{x}u+\theta_{y}v)\right]  dudv$.
Once one obtains the distribution $\left\vert E_{\text{o}}(\theta_{x}%
,\theta_{y},f)\right\vert ^{2}$ from the camera which is located at the focal
plane (see Fig. 1), $(\Delta\theta_{q})^{2}$ can be evaluated through the
following second-order moment \cite{Siegman1990}
\begin{equation}
\left(  \Delta\theta_{q}\right)  ^{2}=\frac{\iint(\theta_{q}-\bar{\theta}%
_{q})^{2}\left\vert E_{\text{o}}(\theta_{x},\theta_{y},f)\right\vert
^{2}d\theta_{x}d\theta_{y}}{\iint\left\vert E_{\text{o}}(\theta_{x},\theta
_{y},f)\right\vert ^{2}d\theta_{x}d\theta_{y}}\text{,} \label{theta_x}%
\end{equation}
where $\bar{\theta}_{q}=\frac{\iint\theta_{q}\left\vert E_{\text{o}}%
(\theta_{x},\theta_{y},f)\right\vert ^{2}d\theta_{x}d\theta_{y}}%
{\iint\left\vert E_{\text{o}}(\theta_{x},\theta_{y},f)\right\vert ^{2}%
d\theta_{x}d\theta_{y}}$ are the average values of transverse spatial
frequencies. In principle the values $(\Delta\theta_{q})^{2}$ are not related
with the focal length\ $f$. However, in order to obtain the good quality of
intensity distribution on the camera, we use a lens with a long focal length
($f=1000$ mm and $300$ mm) in the experiment.

Another quantity to describe vortex beams is the vortex strength, which is a
unique quantity describing the phase structures associated with orbital
angular momentum of FVBs. Since the evolution of FVBs in free space is very
complex and there are infinite pairs of positive and negative vortices
\cite{Berry2004,Leach2004,Lee2004}, we also consider the vortex strength at
the focal plane of a 2-$f$ focusing system. The total vortex strength in
cylindrical coordinates is written as \cite{Berry2004,Jesus2012}%
\begin{align}
S_{\alpha}  &  =\underset{\rho\rightarrow\infty}{\lim}\frac{1}{2\pi}\int
_{0}^{2\pi}d\varphi\arg E_{\text{o,}\alpha}(\rho,\varphi,f)\nonumber\\
&  =\underset{\rho\rightarrow\infty}{\lim}\frac{1}{2\pi}\int_{0}^{2\pi
}d\varphi\text{Re}[iE_{\text{o,}\alpha}^{-1}(\rho,\varphi,f)\partial
E_{\text{o,}\alpha}(\rho,\varphi,f)/\partial\varphi]. \label{vortex strength}%
\end{align}
Here we emphasize that, at the focal plane of a 2-$f$ focusing system, such
FVBs have a finite number of vortices, thus it is more convenient to
investigate the vortex strength of beams at this plane. In order to verify the
vortex strength, we use two experimental methods to identify all vortices
contained in FVBs at the focal plane.

\begin{figure*}[htbp]
\centering
\includegraphics[width=16cm]{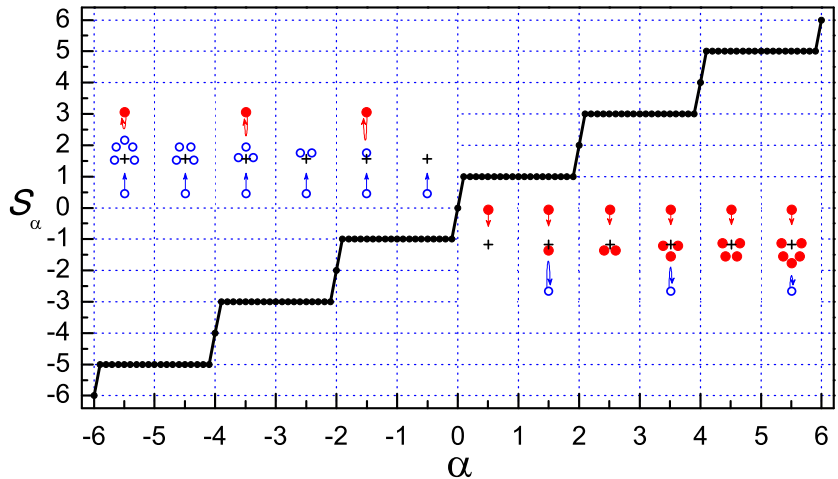}
\caption{(Color) Numerical results of the
vortex strength of FVBs at focal plane as a function of topological charge
$\alpha$ with steps of 0.1. The inset figures are the schematic diagrams of
vortex dynamics for FVBs at focal plane as $\alpha$ increases within each
interval between any two integers, where red-solid and blue-open circles,
respectively, denote the vortices with +1 and -1 charge. The arrows indicate
the movement directions of new vortices, and the cross symbol is the geometric
center on the focal plane.}%
\end{figure*}

\section{Experimental Results and Discussions}

Our main experimental setup is shown in Fig. 1. A SLM (Holoeye
PLUTO-2-VIS-056) is used for reconstruct a FVB, and it is illuminated by a
He-Ne linear-polarized fundamental-mode laser beam with wavelength 632.8 nm.
Here the SLM is applied by grayscale sinusoidal-like holograms, which are
generated first by interfering the wanted first order diffracting beam hosting
the fractional phase azimuthal variation with a plane wave mathematically,
thus the first-order diffraction field in the output field from the SLM is a
FVB. Of course, this SLM can also directly load the patterns of fractional
spiral phase structures, and the output field directly becomes a FVB. The
half-wave plate (HWP) and the beam polarized splitter (PBS) are assembly used
to control the intensity of the incident beam. The output beam from the PBS is
horizontal polarization (along $x$ direction). The expander consists of a pair
of lenses and an aperture, which roughly expands the beam waist to
$w_{0}\approx1$ mm. The SLM is placed within the collimated distance of the
beam after the expander. After modulating by the SLM, the light fields become
the wanted vortex beams with various topological charges, which via a 2-$f$
lens system are carefully recorded by a camera. In our experiment, in order to
reduce the measure error, we chose two different lenses (L$_{1}$) with $f=300$
mm and $f=1000$ mm to repeatedly measure the intensity distributions near or
at their focal planes, and meanwhile the 2-$f$ lens system should be precisely
adjusted by keeping both the distances from the SLM to L$_{1}$ and from
L$_{1}$ to the camera equal to $f$. The camera sensor (Sony IMX183(M)) has a
$13.06$ mm $\times8.76$ mm exposure area with each pixel size of $2.4$ $\mu$m
$\times2.4$ $\mu$m and real-time 12-bit data depth.

In the inset of Fig. 1, it shows the intensity distributions of FVBs with
$\alpha=1.5$ and $3.5$, respectively, locating before, at, and after the
L$_{1}$'s focal plane. When the position of the camera is at the focal plane,
the intensity profile is exactly symmetric about $y$ axis and all vortices (or
called as singularities) in light fields are aligned (or symmetric) with $y$
axis; while when it is away from the focal plane, the intensity profile is not
symmetric any more and noncentral vortices rotate around the geometric center
of $x$-$y$ plane. This property enables us to accurately measure the intensity
distributions at focal plane of such a 2-$f$ lens system. According to the
formula in the previous section, we can acquire the information of both
$|E_{\text{o}}(x,y,f)|^{2}$ and $|E_{\text{o}}(\theta_{x},\theta_{y},f)|^{2}$.
Note that we should consider both the size of the pixels on the camera and the
value of $f $ to get $|E_{\text{o}}(\theta_{x},\theta_{y},f)|^{2}$. Meanwhile,
we also record the input intensity distributions $|E_{\text{i}}(u,v,0)|^{2}$
at the position of the SLM, and these input intensities are also repeatedly
measured for better determining practical values of $w_{x}$ and $w_{y}$. From
$|E_{\text{o}}(\theta_{x},\theta_{y},f)|^{2}$, one can know the values of
$(\Delta\theta_{q})^{2}$.

Before we discuss our experimental results, in Fig. 2, let us first see the
numerical result on total vortex strength $S_{\alpha}$ of such FVBs as a
function of topological charge $\alpha$. Different from the previous results
\cite{Berry2004,Jesus2012}, here we observe the jumps in $S_{\alpha} $ only
when $\alpha$ is around any even number (i.e., $\alpha=0,\pm2,\pm4,\pm
6,\cdots$). Actually there are two continuous jumps around each even integer
number: the first jump is that $S_{\alpha}$ increases 1 when $\alpha$
approaches to any even integer, and $S_{\alpha}$ exactly equals to that even
integer when $\alpha$ is an even integer, and the second jump is that
$S_{\alpha}$ continues to increase 1 when $\alpha$ is slightly larger than any
even number. However, in Ref. \cite{Berry2004}, the jumps in $S_{\alpha}$
happen at every half-integer value of $\alpha$, and $S_{\alpha}$ only
increases 1 at each jump for indicating a birth of a vortex; and in Ref.
\cite{Jesus2012}, each jump leads to a unit change in $S_{\alpha}$ only at
$\alpha=n+\varepsilon$, where $n$ is an integer and $\varepsilon$ is a small
fraction. From Fig. 2, it can be concluded that
\begin{subequations}
\[
S_{\alpha}=\left\{
\begin{array}
[c]{c}%
m,\text{ \ \ \ \ \ \ \ \ \ for }\alpha=m\text{\ \ }\\
2m-1,\text{ for }2(m-1)<\alpha<2m
\end{array}
\right.
\]
where $m$ is integer. Here we would like to point out that the jumps in
$S_{\alpha}$ here do not completely indicate the birth of a vortex. These
features will be explained next by phase structures and the dynamics of
vortices at focal plane.

\begin{figure*}[t]
\centering
\includegraphics[width=17cm]{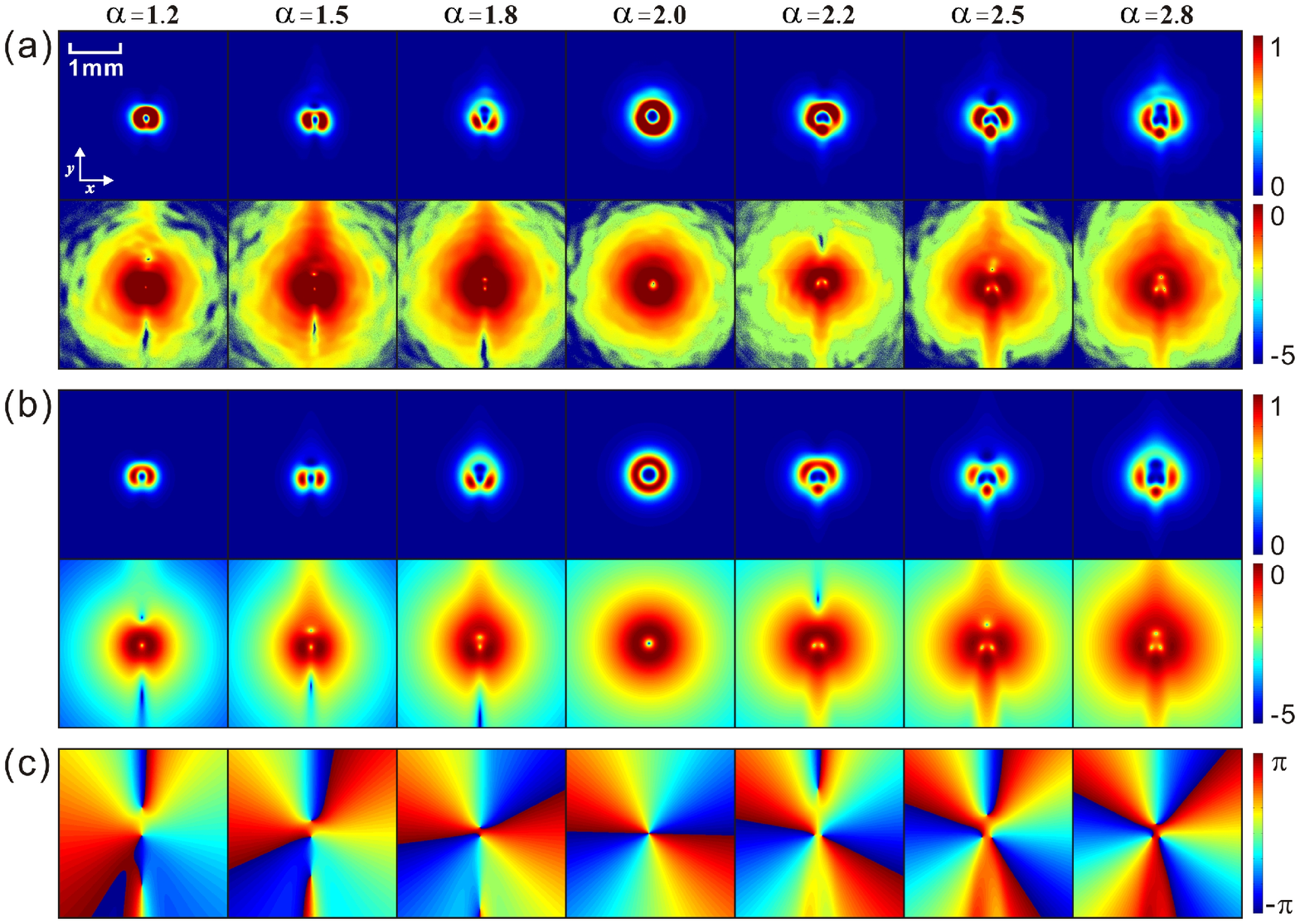}
\newline\caption{(Color) (a) Experimental
and (b) numerical results of typical intensity distributions for FVBs with
different values $\alpha$ from 1.2 (left) to 2.8 (right). The corresponding
numerical phase structures are plotted in (c). In (a) and (b), the first and
second rows are, respectively, in normal and logarithmic scales. Here we use
the lens L$_{1}$ with $f=1000$ mm. }%
\end{figure*}

Figure 3 (a) experimentally shows typical intensity distributions of FVBs with
different values $\alpha$ at focal plane. In order to prominently display the
singularities in low-intensity regions of light fields, the camera measuring
intensities works at overexposed mode for the second row of Fig. 3(a). It is
seen that there are three deep-dark regions for $\alpha=1.2$, $1.5$, and
$1.8$. The position of the near-center dark region (this vortex with $+1$
charge) only suffers little change due to the movement of a pair of top and
bottom vortices. From $\alpha=1.2$ to $1.5$ in Fig. 3(a), it is observed
that\ both the top and bottom vortices, with $+1$ and $-1$ charges,
respectively, approach to the center vortex, but the top vortex with $+1$
charge moves faster than the bottom one; while from $\alpha=1.5$ to $1.8$, the
top vortex continues to close the near-center vortex but the bottom vortex
quickly leaves away from the near-center vortex, also see the schematic
dynamics of vortices in the inset of Fig. 2. When $\alpha=2$, the top vortex
completely merges with the center vortex, and the bottom vortex moves to
infinity (or say, it disappears). Thus there is a jump in $S_{\alpha}$ in Fig.
2 when $\alpha$ closes to 2, because a vortex with $-1$ charge disappears.

For $\alpha=2.2$ to $2.8$ in Fig. 3, one can observe that there are still
three-dark regions (singularities): two vortices locate near the center and
move very slightly, and other top vortex moves quickly from the outside to
center. In fact, all these three vortices have the same $+1$ charge
(experimentally verified below). Therefore, when $\alpha$ is slightly larger
than 2, in principle a new vortex is born from outside infinity and it moves
to the center as $\alpha$ closes to 3, also see the schematic dynamics in
inset of Fig. 2. Experimental results in Fig. 3(a) are in good agreement with
the corresponding numerical results in Fig. 3(b). For further demonstration on
evolutions of these singularities (or vortices), the individual phase
structures are also plotted in Fig. 3(c). Clearly, there occur a pair of
positive and negative charge vortices for $\alpha\in(1,2)$, which leads
$S_{\alpha}$ unchanged. Meanwhile a new +1 vortex is generated for $\alpha
\in(2,3)$. From Fig. 2 and Fig. 3, it is concluded that, when $2(m-1)<\alpha
<2m-1$, $S_{\alpha}=2m-1$ and there is a new vortex moving from outside to
center; when $2m-1<\alpha<2m$, $S_{\alpha}\ $still equals to $2m-1$ but a pair
of +1 and -1 charge vortices are generated. The layouts of all vortices for
different FVBs at focal plane are also schematically plotted in Fig. 2. This
explains the appearance of the two continuous jumps in $S_{\alpha}$ around
each even number of $\alpha$.

\begin{figure*}[htbp]
\centering
\includegraphics[width=16.5cm]{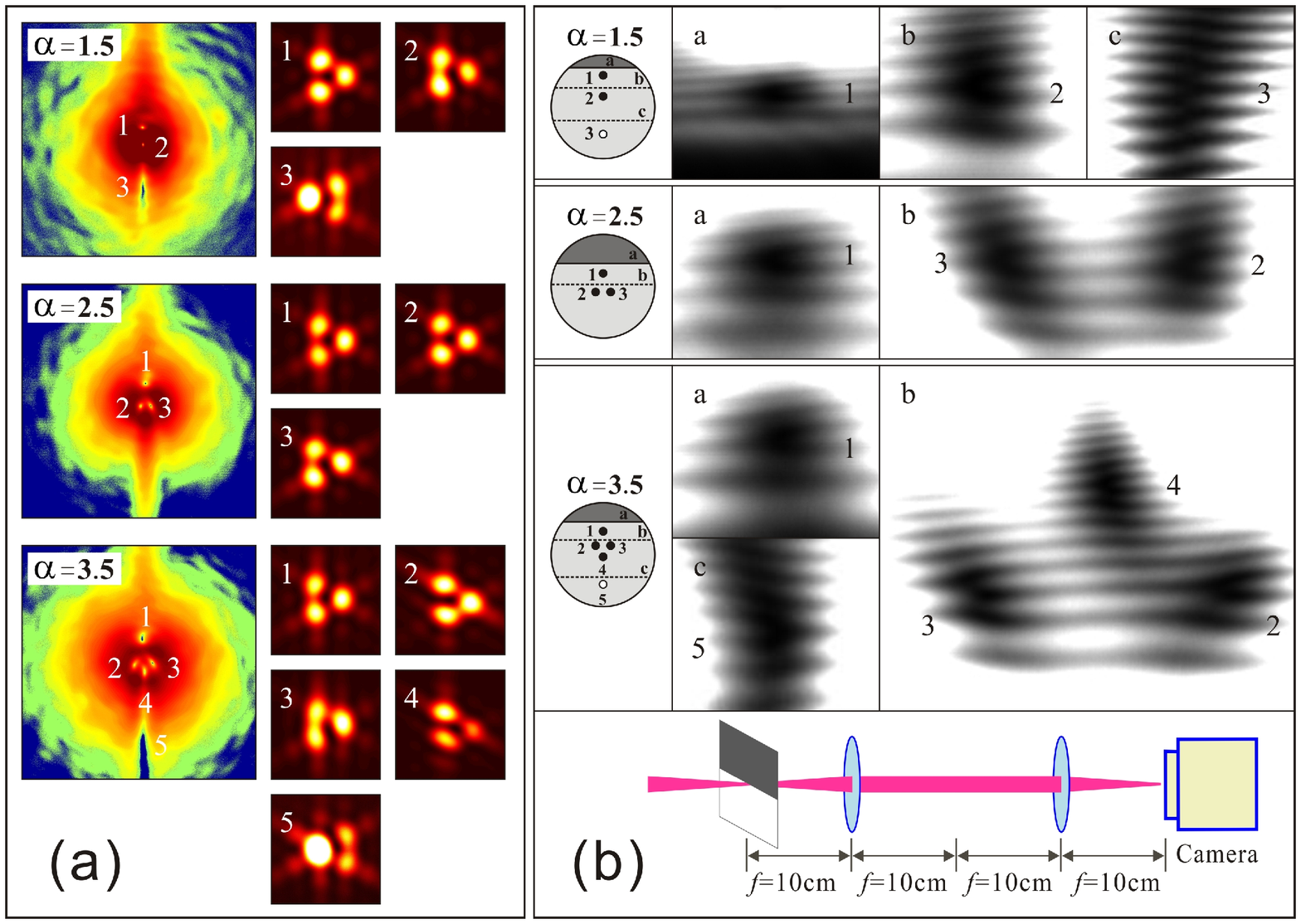}
\caption{(Color) Experimental
verification on the topological charge of each vortex in FVBs at the focal
plane of a 2-$f$ lens system. (a) The vortex distributions in different FVBs
(left), and the corresponding interference patterns (right) of individual
vortices by the diffraction method of a tiny triangle aperture. (b) The
interference patterns diffracted by the straight blade via a 4-$f$ lens
system. Here the blade edge is moved from up to down, and is located at
different positions a, b, or c [see the left sketch of each case in (b)], for
detecting the fork-like interference patterns. The measurement system by the
straight blade is also plotted in the bottom of (b). Each vortex and its
interference patten in (a) and (b) are correspondingly marked with a number,
and here the topological charge $\alpha=1.5, 2.5, 3.5.$ }%
\end{figure*}

Then we ask ourselves that, are all these vortices in FVBs measurable? In
order to solidly confirm our observations, we use two methods to confirm our
findings in Fig. 2 and Fig. 3. In Fig. 4(a), it demonstrates the experimental
verification of these vortices in FVBs by using a tiny equilateral triangle
aperture (confirmed under an microscope). This kind of method for testing the
vortex was proposed by \cite{Hickmann2010} and has been widely used to detect
vortex due to its high efficiency and accuracy
\cite{Jesus2012,Mourka2011,Anderson2012}. Here the triangle aperture is placed
on a two-dimensional movable stage, locating at the L$_{1}$'s focal plane via
the reflection of the BS. This measurement optical system is shown in Fig. 1,
and another 2-$f$ system of the lens L$_{2}$ is used to collect interference
pattern at its focal plane. From Fig. 4(a), for $\alpha=1.5$, three vortices
in intensity are detected and verified. It is seen that the 1st and 2nd
interference patterns for both the top and near-center vortices are the same
and indicate the vortices with +1 charge, but the 3rd pattern for the bottom
vortex is rotated by $\pi$ and it indicates this vortex with $-1$ charge. For
$\alpha=2.5 $, it is not difficult to observe each interference pattern for
each vortex and all these patterns indicate the vortices having the same sign
with +1 charge. For $\alpha=3.5$, there are similar patterns for the top and
three near-center vortices with labels 1 to 4, indicating the same sign with
+1 charge, but the last pattern for the bottom vortex with label 5 is also
rotated by $\pi$ and it tells us that it is a negative vortex with -1 charge.
It should be noted that one has to increase the input power of laser in order
to measure the bottom vortices in cases of $\alpha=1.5$ and $3.5$ because
their intensities are usually very weak.

Another verification measurement system is shown in the bottom of Fig. 4(b).
It consists of two lenses forming a 4-$f$ optical system, which can inversely
image the pattern located at its input plane. However, due to the diffraction
from the edge of the straight blade, when the blade edge approaches a vortex
contained in light fields, one can observe a fork-like interference pattern at
the image plane. Via zooming in interference patterns, it is seen that there
are fork-like interference fringes, which are formed from the interference
between an optical vortex contained in light fields and an edge diffraction
wave by the blade edge. The open direction of the fork indicates the sign of a
vortex. For example, when $\alpha=1.5$, one can see that the 1st and 2nd fork
interference fringes are similar and indicate the vortices with +1 charge at
the fork centers, and the 3rd fork fringes have opposite directions, which
denote a vortex with -1 charge existing at its fork center. Similarly, one can
use this method to determine all vortices for the cases of $\alpha=2.5$ and
$3.5$ by moving the blade edge. In a word, this method again confirms that a
pair of positive and negative vortices exist for $\alpha=1.5$ and $3.5$, and
all vortices have the same sign for $\alpha=2.5$. By this method, one can
verify other situations for different $\alpha$. Therefore we conclude that
total $S_{\alpha}$ for FVBs follows Fig. 2, which is different from Refs.
\cite{Berry2004,Jesus2012}.

\begin{figure}[htbp]
\centering
\includegraphics[width=8.5cm]{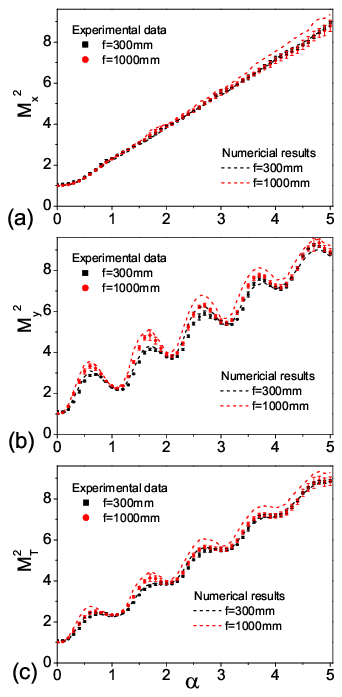}
\caption{(Color) Experimental results
(black and red dots), and matched numerical results (black and red dashed
lines) of the BPF values in (a) $x$-direction, (b) $y$-direction, and (c)
total BPF values of FVBs as functions of $\alpha$ with steps of 0.1. The error
bars are obtained from multiple measurements and the focal length of the lens
L$_{1}$ is chosen to be $f=300$ mm and $f=1000$ mm for two groups of multiple
measurements. }%
\end{figure}

As we presented in the section II, the distributions of light fields at focal
plane represent the behavior of light at far-field regions. In Refs.
\cite{Leach2004}, one already knew that there are rotation processes of vortex
during the propagations of FVBs in free space, and both birth and annihilation
phenomena of vortices may happen as $\alpha$ increases. From above
discussions, when $\alpha$ increases, the vortices in FVBs show different
traces moving on focal plane. Hence, it is very meaningful to evaluate the BPF
of such FVBs, which is never investigated before. Figure 5 shows the measured
results of BPF values as a function of topological charge $\alpha$. It is
observed in Fig. 5(a) and 5(b) that as $\alpha$ increases, the BPF $M_{x}^{2}$
in $x$-direction almost increases linearly, while $M_{y}^{2}$ in $y$-direction
increases with oscillations. Different behaviors of $M_{x}^{2}$ and $M_{y}%
^{2}$ can be understood by intensity distributions at focal plane from Fig. 3.
The intensity distributions for such beams are always symmetric in $x$
direction at focal plane or at far-field region of free space, and the
effective diverging angle (also corresponding to the effective spread of
spatial frequency) almost increases linearly too as $\alpha$ increases.
However, in $y$ direction, a vortex or a pair of vortices at focal plane can
be born periodically with the increase of $\alpha$, and their movements (see
the sketch in Fig. 2) along $y$ direction lead to the oscillations in
$M_{y}^{2}$. The numerical curves are obtained by substituting Eq.
(\ref{Expansion22}) into Eq. (\ref{theta_x}) and using Eqs. (\ref{M2x}) and
(\ref{M2y}), and the numerical integration range in Eq. (\ref{theta_x}) should
be matched with experimental data, because the numerator integration of
Eq.(10) depends on its integrating range, and in practical sense, there are
also the physical limitation of camera camera, which cannot record the
intensity values lower than its threshold. It is seen that, when the numerical
ranges are matched with the experimental data, the measured BPF values are in
good agreement with the matched numerical results. When we average both
$M_{x}^{2}$ and $M_{y}^{2}$ to define the total BPF $M_{T}^{2}=\frac{1}%
{2}(M_{x}^{2}+M_{y}^{2})$, then there are step-like jumps in $M_{T}^{2}$ as
$\alpha$ increases, see Fig. 5(c). This curve has certain similarity to the
practical changes of the effective OAM of such beams
\cite{Alperin2016,Alperin2017}. For only integer values of $\alpha$,
$M_{T}^{2}$ has similar property of linear increase, like those
Laguerre-Gaussian beams as a function of integer azimuthal index $n$
\cite{Saghafi1998,Vega2007}. 

\section{Summary}

In conclusion, we have studied the vortex strength and BPF of FVBs. Under
paraxial approximation, we have derived the output filed for such FVBs passing
through arbitrary linear ABCD optical system. From light fields at focal plane
of a 2-$f$ lens system (similar to the far-field distribution), our results
have demonstrated the vortex strength increases by unit continuously when
topological charge is very close before and after an even number. Triangle
aperture and straight blade, respectively, were employed to experimentally
detect the vortices of FVBs at focal plane and to confirm each vortex charge,
and thus the change of vortex strength for FVBs is confirmed. The experimental
results, demonstrating the movement of all vortices at focal plane, are in
good agreement with the numerical results, and the dynamics of all vortices at
focal plane are different in two intervals of topological charges. The BPF
value has been used to describe the characteristics of propagation behavior of
FVBs. Experimental results show that as the topological charge increases, the
value of BPF increases linearly in $x$-component while the obvious
oscillations exist in the BPF value in $y$-component. Moreover, there are
step-like jumps in the total BPF values as the topological charge increases,
which have a similar behavior to the OAM of FVBs as the topological charge
increases. Our findings will bring distinct perspectives to the research of
vortex strength at focal plane and the BPF value for both FVBs and other
complex structured light fields.
\end{subequations}
\begin{acknowledgments}
Zhejiang Provincial Natural Science Foundation of China under Grant No.
LD18A040001; National Key Research and Development Program of China (No.
2017YFA0304202); National Natural Science Foundation of China(grants No.
11674284 and U1330203); Fundamental Research Funds for the Center Universities
(No. 2017FZA3005).
\end{acknowledgments}


\begin{thebibliography}{99}                                                                                               %


\bibitem {Allen1992}L. Allen, M. W. Beijersbergen, R. J. C. Spreeuw, and J. P.
Woerdman, Orbital angular momentum of light and the transformation of
Laguerre-Gaussian laser modes, Phy. Rev. A \textbf{45}, 8185 (1992).

\bibitem {Berry2004}M. V. Berry, Optical vortices evolving from helicoidal
integer and fractional phase steps, J. Opt. A: Pure Appl. \textbf{6}, 259 (2004).

\bibitem {Leach2004}J. Leach, E. Yao, and M. J. Padgett, Observation of the
vortex structure of a non-integer vortex beam, New J. Phys. \textbf{6}, 71 (2004).

\bibitem {Alperin2017}S. N. Alperin, and M. E. Siemens, Angular momentum of
topologically structured darkness, Phy. Rev. Lett. \textbf{119}, 203902 (2017).

\bibitem {Georgiy2017}G. Tkachenko, M. Z. Chen, K. Dholakia, and M. Mazilu, Is
it possible to create a perfect fractional vortex beam?, Optica \textbf{4},
330 (2017).

\bibitem {Tao2005}S. H. Tao, X. C. Yuan, J. Lin, X. Peng, and H. B. Niu,
Fractional optical vortex beam induced rotation of particles, Opt. Express
\textbf{13}, 7726 (2005).

\bibitem {Gbur2016}G. Gbur, Fractional vortex Hilbert's hotel, Optica
\textbf{3}, 222 (2016).

\bibitem {Hong2015}Z. Y. Hong, J. Zhang, and B. W. Drinkwater, On the
radiation force fields of fractional-order acoustic vortices, EPL
\textbf{110}, 14002 (2015).

\bibitem {Jia2018}Y. R. Jia, Q. Wei, D. J. Wu, Z. Xu, and X. J. Liu,
Generation of fractional acoustic vortex with a discrete Archimedean spiral
structure plate, Appl. Phys. Lett. \textbf{112}, 173501 (2018).

\bibitem {Bandyopadhyay2017}P. Bandyopadhyay, B. Basu, and D. Chowdhury,
Geometric phase and fractional orbital-angular-momentum states in electron
vortex beams, Phy. Rev. A \textbf{95}, 013821 (2017).

\bibitem {Oemrawsingh2004}S. S. R. Oemrawsingh, A. Aiello, E. R. Eliel, G.
Nienhuis, and J. P. Woerdman, How to observe high-dimensional two-photon
entanglement with only two detectors, Phy. Rev. Lett. \textbf{92}, 217901 (2004).

\bibitem {Oemrawsingh2005}S. S. R. Oemrawsingh, X. Ma, D. Voigt, A. Aiello, E.
R. Eliel, G. W. 't Hooft, and J. P. Woerdman, Experimental demonstration of
fractional orbital angular momentum entanglement of two photons, Phy. Rev.
Lett. \textbf{95}, 240501 (2005).

\bibitem {Chen2014}L. X. Chen, J. J. Lei, and J. Romero, Quantum digital
spiral imaging, Light: Sci. Appl. \textbf{3}, e153 (2014).

\bibitem {Basistiy1995}I. V. Basistiy, M. S. Soskin, and M. V. Vasnetsov,
Optical wavefront dislocations and their properties, Opt. Commun.
\textbf{119}, 604 (1995).

\bibitem {Basistiy2004}I. V. Basistiy, V. A. Pas'ko, V. V. Slyusar, M. S.
Soskin, and M. V. Vasnetsov, Synthesis and analysis of optical vortices with
fractional topological charges, J. Opt. \textbf{6}, S166 (2004).

\bibitem {Alperin2016}S. N. Alperin, R. D. Niederriter, J. T. Gopinath, and M.
E. Siemens, Quantitative measurement of the orbital angular momentum of light
with a single, stationary lens, Opt. Lett. \textbf{41}, 5019 (2016).

\bibitem {Lee2004}W. M. Lee, X. C. Yuan, and K. Dholakia, Experimental
observation of optical vortex evolution in a Gaussian beam with an embedded
fractional phase step, Opt. Commun. \textbf{239}, 129 (2004).

\bibitem {Fang2017}Y. Q. Fang, Q. H. Lu, X. L. Wang, W. H. Zhang, and L. X.
Chen, Fractional-topological-charge-induced vortex birth and splitting of
light fields on the submicron scale, Phy. Rev. A \textbf{95}, 023821 (2017).

\bibitem {Jesus2012}A. J. Jesus-Silva, E. J. S. Fonseca, and J. M. Hickmann,
Study of the birth of a vortex at Fraunhofer zone, Opt. Lett. \textbf{37},
4552 (2012).

\bibitem {Siegman1990}A. E. Siegman, New developments in laser resonators,
Proc. SPIE \textbf{1224}, 2 (1990).

\bibitem {Borghi1997}R. Borghi and M. Santarsiero, $M^{2}$ factor of
Bessel-Gauss beams, Opt. Lett. \textbf{22}, 262 (1997).

\bibitem {Porras2001}M. A. Porras, R. Borghi, and M. Santarsiero, Relationship
between elegant Laguerre-Gauss and Bessel-Gauss beams, J. Opt. Soc. Am. A
\textbf{18}, 177 (2001).

\bibitem {Saghafi1998}S. Saghafi, and C. J. R. Sheppard, The beam propagation
factor for higher order Gaussian beams, Opt. commun. \textbf{153}, 207 (1998).

\bibitem {Vega2007}J. C. Guti\'{e}rrez-Vega, Fractionalization of optical
beams: II. Elegant Laguerre-Gaussian modes, Opt. Express \textbf{15}, 6300 (2007).

\bibitem {Ramee2000}S. Ramee and R. Simon, Effect of holes and vortices on
beam quality, J. Opt. Soc. Am. A \textbf{17}, 84 (2000).

\bibitem {Collins1970}S. A. Collins, Lens-system diffraction integral written
in terms of matrix optics, J. Opt. Soc. Am. \textbf{60}, 1168 (1970).

\bibitem {Zhao2000}S. Wang, and D. Zhao, \emph{Matrix Optics}, (Springer,
Berlin, 2000).

\bibitem {Hickmann2010}J. M. Hickmann, E. J. S. Fonseca, W. C. Soares, and S.
Ch\'{a}vez-Cerda, Unveiling a truncated optical lattice associated with a
triangular aperture using light's orbital angular momentum, Phys. Rev. Lett.
\textbf{105}, 053904 (2010).

\bibitem {Mourka2011}A. Mourka, J. Baumgartl, C. Shanor, K. Dholakia, and E.
M. Wright, Visualization of the birth of an optical vortex using diffraction
from a triangular aperture, Opt. Express \textbf{19}, 5760 (2011).

\bibitem {Anderson2012}M. E. Anderson, H. Bigman, L. E. E. de Araujo, and J.
L. Chaloupka, Measuring the topological charge of ultrabroadband,
optical-vortex beams with a triangular aperture, J. Opt. Soc. Am. B
\textbf{29}, 1968 (2012).
\end{thebibliography}
\end{document}